\documentclass[12pt]{iopart}
\usepackage{iopams}
\usepackage{epsf,graphics,graphicx}
\usepackage{float}
\usepackage{subfigure}
\newcommand{\ket}[1]{|#1\rangle}
\newcommand{\bra}[1]{\langle #1|}

\begin{document}
\title{Non-adiabatic holonomic quantum computation}
\author{Erik Sj\"oqvist$^{1,2}$}
\ead{erik.sjoqvist@kvac.uu.se}
\author{D. M. Tong$^3$}
\ead{tdm@sdu.edu.cn}
\author{L. Mauritz Andersson$^4$}
\ead{mauritza@kth.se}
\author{Bj\"orn Hessmo$^{1}$}
\ead{phyhbg@nus.edu.sg}
\author{Markus Johansson$^{1,2}$}
\ead{cqtbemj@nus.edu.sg}
\author{Kuldip Singh$^{1}$}
\ead{sciks@nus.edu.sg}
\vskip 0.5 cm
\address{$^1$Centre for Quantum Technologies, National University of Singapore,
3 Science Drive 2, 117543 Singapore, Singapore}
\address{$^2$Department of Quantum Chemistry, Uppsala University, Box 518,
SE-751 20 Uppsala, Sweden, EU}
\address{$^3$Physics Department, Shandong University, Jinan, 250100, China}
\address{$^4$Department of Applied Physics, KTH Royal Institute of Technology, 
SE-100 44 Stockholm, Sweden, EU}
\date{\today}
\begin{abstract}
We develop a non-adiabatic generalization of holonomic quantum computation in which high-speed
universal quantum gates can be realized by using non-Abelian geometric phases. We show how
a set of non-adiabatic holonomic one- and two-qubit gates can be implemented by utilizing
optical transitions in a generic three-level $\Lambda$ configuration. Our scheme opens up
for universal holonomic quantum computation on qubits characterized by short coherence time.
\end{abstract}
\pacs{03.65.Vf, 03.67.Lx}
\submitto{\NJP}
\maketitle
\section{Introduction}
Circuit-based quantum computation relies on the ability to perform a universal
set of quantum gate operations on a set of quantum-mechanical bits (qubits). A key
challenge to achieve this goal is to find implementations of gates that are resilient to
certain kinds of errors. Holonomic quantum computation (HQC) \cite{zanardi99} is a
general procedure to build universal sets of robust gates, by using non-Abelian geometric
phases \cite{wilczek84}.

HQC is conventionally based on adiabatic evolution. The idea is to encode a set of qubits in 
a set of degenerate eigenstates of a parameter dependent Hamiltonian and to adiabatically 
transport these states around a loop in the corresponding parameter space. This effectuates 
a holonomic gate acting on the qubits. It has been shown \cite{zanardi99} that adiabatic 
quantum holonomies generically allows for universal quantum computation. 

Adiabatic holonomic gates have been proposed for trapped ions \cite{duan01}, superconducting 
nanocircuits \cite{faoro03} and semiconductor quantum dots \cite{solinas03a}. These gates still 
await experimental realization. An obstacle in achieving this is the long run-time required for 
the desired parametric control associated with adiabatic evolution. In other words, as these 
gates operate slowly compared to the dynamical time scale, they become vulnerable to open 
system effects and parameter fluctuations that may lead to loss of coherence. On the other 
hand, if the run-time is decreased in order to shorten the exposure, non-adiabatic corrections 
start to become significant and the parametric control is lost. These problems have been tackled 
\cite{wang01} by using Abelian non-adiabatic geometric phases \cite{aharonov87} to realize 
quantum gates. However, such geometric phase gates are limited to commuting operations and 
thus cannot perform universal holonomic quantum computation.

To combine speed and universality, we propose here a generalization of HQC based on 
non-adiabatic non-Abelian geometric phases proposed in \cite{anandan88}. The key advantage 
of our holonomic setting is that it removes the problem of long run-time associated with the original 
form of HQC \cite{zanardi99}. We demonstrate an experimentally feasible optical scheme to implement 
a universal set of holonomic one- and two-qubit gates for non-adiabatic optical transitions in 
three-level $\Lambda$ configurations. The proposed setup allows for any quantum computation 
on any number of qubits by purely geometric means. 

The outline of the paper is as follows. The general theory of non-adiabatic HQC is described 
in the next section. In section \ref{sec:implementation}, we demonstrate a universal set of 
non-adiabatic holonomic gates in a generic $\Lambda$ configuration and show that these  
gates can be made robust to decay. The non-adiabatic holonomic gates are interpreted 
geometrically in section \ref{sec:geometry}. The paper ends with the conclusions.  

\section{Non-adiabatic holonomic quantum computation} 
\label{sec:theory}
Consider a quantum system characterized by an $N$-dimensional Hilbert space. A 
computational system, typically a set of qubits, is encoded in a $K$-dimensional subspace 
$M(0)$ of Hilbert space. A quantum gate that manipulates the computational state can be 
induced by taking $M(0)$ around a smooth path $C: [0,\tau] \ni t\mapsto M (t)$ of 
$K$-dimensional subspaces in such a way that $M(\tau) = M(0)$. Thus, $C$ is a loop of 
such subspaces generated by a suitable Hamiltonian $H(t)$ of the full system. In this way, 
any computational state residing in $M(0)$ will in general end up in a new state in the same 
subspace. The unitary transformation relating the final and initial states is the quantum gate. 
The idea of non-adiabatic holonomic quantum computation is to make the resulting gate $C$ 
dependent, but independent of any dynamical parameters such as the run time $\tau$ and the 
energies of the system. 

Let us formalize this idea by introducing a once differentiable set of orthonormal ordered 
bases $\ket{\zeta_k (t)}$, $k=1,\ldots ,K$, of $M(t)$ along $C$, such that $\ket{\zeta_k (\tau)} = 
\ket{\zeta_k (0)}$. One may vizualise $\ket{\zeta_k (t)}$ as a $K$-tuple of vectors moving in the 
$N$-dimensional Hilbert of the full system. The final time evolution operator projected onto the 
initial subspace may be written as ($\hbar = 1$ from now on) \cite{anandan88}
\begin{eqnarray}
U(\tau,0) =
\sum_{k,l=1}^K \left( {\bf T} e^{i\int_0^{\tau} (\boldsymbol{A} (t) -
\boldsymbol{H} (t)) dt} \right)_{kl} \ket{\zeta_k (0)} \bra{\zeta_l (0)},
\end{eqnarray}
where ${\bf T}$ is time ordering. Here, $\boldsymbol{A}_{kl} (t) = i\bra{\zeta_k (t)}
\dot{\zeta}_l (t) \rangle$ and $\boldsymbol{H}_{kl} (t) = \bra{\zeta_k (t)}
H(t) \ket{\zeta_l (t)}$ are Hermitian $K\times K$ matrices.  Thus, $U(\tau,0)$ is a 
unitary operator on $M(0)$. 

To understand the meaning of $\boldsymbol{A}$, let us check how it transforms 
under a smooth change of basis spanning $M(t)$. Such a transformation is known as a 
gauge transformation as it changes the basis but not the subspace itself.  Explicitly, 
if $\ket{\zeta_k (t)} \rightarrow \sum_{l=1}^K \ket{\zeta_l (t)} \boldsymbol{V}_{lk} (t)$, 
$\boldsymbol{V} (t)$ being a once differentiable family of unitary $K \times K$ matrices 
such that $\boldsymbol{V} (\tau) = \boldsymbol{V} (0)$, then  $\boldsymbol{A} \rightarrow 
\boldsymbol{V}^{\dagger} \boldsymbol{A}  \boldsymbol{V} + i\boldsymbol{V}^{\dagger} 
\dot{\boldsymbol{V}}$. This shows that $\boldsymbol{A}$ transforms as a proper vector 
potential. Thus,  the unitary
\begin{eqnarray}
\boldsymbol{U} = {\bf P} e^{i\oint_{C} \boldsymbol{\mathcal{A}}} , 
\label{eq:general}
\end{eqnarray}
$\boldsymbol{\mathcal{A}}_{kl} = i\bra{\zeta_k (t)} d \zeta_l (t) \rangle$ being the 
matrix-valued connection one-form, is the holonomy matrix generalizing the Wilczek-Zee 
holonomy \cite{wilczek84} to non-adiabatic evolutions. Note that ${\bf P}$ is path ordering along 
$C$ and that $\boldsymbol{U} \rightarrow \boldsymbol{V}^{\dagger} (0) \boldsymbol{U} 
\boldsymbol{V} (0)$ under a gauge transformation. This gauge covariance essentially means 
that the holonomy matrix is a property of the loop $C$ and we may write 
$\boldsymbol{U} \equiv \boldsymbol{U}(C)$.

The following two conditions are necessary for universal non-adiabatic HQC: (i) there should 
exist physically accessible loops $C$ of subspaces along which the Hamiltonian matrix 
$\boldsymbol{H}_{kl} (t) = \bra{\zeta_k (t)} H(t) \ket{\zeta_l (t)}$ vanishes; (ii) there should
exist at least two such loops $C$ and $C'$, both based at $M(0)$, for which the corresponding 
$\boldsymbol{U}(C)$ and $\boldsymbol{U}(C')$ do not commute. While the first condition 
assures that the evolution is purely geometric, the second one is necessary to realize 
universality. Under conditions (i) and (ii), there is a set of quantum gates
\begin{eqnarray}
U(\tau,0) = U(C) =
\sum_{k,l = 1}^K \boldsymbol{U}_{kl} (C) \ket{\zeta_k (0)} \bra{\zeta_l (0)}
\end{eqnarray}
that may be able to perform any computation on qubits encoded in $M(0)$ based purely on 
the geometric properties of the subspace paths. We demonstrate that these conditions can be 
met in a generic three-level $\Lambda$ configuration, by means of which a universal set of 
one- and two-qubit gates can be realized.

\section{Physical implementation}
\label{sec:implementation}

\begin{figure}[ht!]
\begin{center}
\includegraphics[width=14 cm]{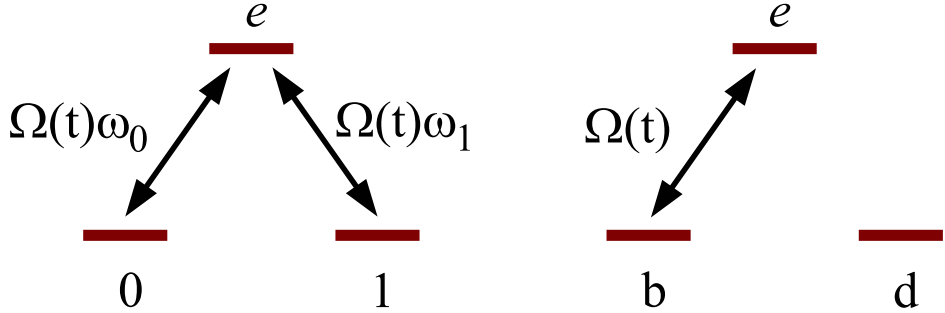}
\end{center}
\caption{\label{fig:1gate} Setup for non-adiabatic holonomic one-qubit gate in a $\Lambda$ 
configuration. A pair of zero-detuned laser pulses couple two ground state levels $0$ and 
$1$ to an excited state $e$ (left panel). The two ground state levels define a single qubit 
and the laser parameters satisfy $|\omega_0|^2 + |\omega_1|^2 = 1$. Note that 
the `bare' ground state levels may be degenerate or non-degenerate since the lasers 
are assumed to be tunable in an independent fashion. The dark state
$\ket{d} = -\omega_1 \ket{0} + \omega_0 \ket{1}$ is decoupled from the bright state
$\ket{b} = \omega_0^{\ast} \ket{0} + \omega_1^{\ast} \ket{1}$ by choosing time-independent
$\omega_0$ and $\omega_1$ over the duration of the pulse pair. The system thereby
performs Rabi oscillations between the bright and excited states with
frequency $\Omega (t)$ (right panel). The evolution of the qubit subspace is purely geometric
and becomes cyclic after completing a Rabi oscillation by choosing $\Omega (t)$ to be a
real-valued $\pi$ pulse. The resulting unitary quantum gate operation acting on the qubit
is determined by the holonomy of the loop traced out by the subspace spanned by 
$e^{-i\int_0^t H^{(1)} (t') dt'} \ket{k}$, $k=0,1$. By applying sequentially two $\pi$ pulse 
pairs with negligible temporal overlap, any desired holonomic one-qubit gate can be realized.}
\end{figure}

\subsection{One-qubit gate}
Consider a three-level atom or ion consisting of the `bare' energy eigenstates $\ket{0},\ket{1}$ 
and $\ket{e}$ with energies $w_0,w_1$ and $w_e$, respectively. These states form a $\Lambda$ 
configuration in which each $k\leftrightarrow e$ transition ($k=0,1$) is driven separately 
by a suitably polarized laser pulse with frequency $\nu_k$. In the rotating frame, the Hamiltonian 
describing the system-laser interaction takes the form 
\begin{eqnarray}
H(t) = \Delta_0 \ket{0} \bra{0} + \Delta_1 \ket{1} \bra{1} + \Omega(t) \left( \omega_0 \ket{e} \bra{0} +
\omega_1 \ket{e} \bra{1} + {\textrm{h.c.}} \right) .
\end{eqnarray}
where we have neglected rapidly oscillating counter-rotating terms (rotating wave approximation). 
Here, $\Delta_k = 2\pi \nu_k - \omega_{ek}$, where $\omega_{ek} = w_e - w_k$, are detunings 
that can be varied independently by changing $\nu_k$. The laser parameters $\omega_0$ 
and $\omega_1$ satisfy $|\omega_0|^2 + |\omega_1|^2 = 1$, and describe the relative strength 
and relative phase of the $0\leftrightarrow e$ and $1 \leftrightarrow e$ transitions. The Hamiltonian 
is turned on and off during the time interval $[0,\tau]$, controlled by the pulse envelope $\Omega (t)$. 
We take $\ket{0}$ and $\ket{1}$ to define the one-qubit state space $M(0)$. 
  
A universal holonomic one-qubit gate can be realized in the above $\Lambda$ system by choosing 
time-independent $\omega_0$ and $\omega_1$ over the duration of the pulse pair and by tuning 
the laser frequencies so that the detunings $\Delta_0$ and $\Delta_1$ vanish. Under these conditions, 
the Hamiltonian reduces to  
\begin{eqnarray}
H^{(1)} (t) & = & \Omega(t) \left( \omega_0 \ket{e} \bra{0} +
\omega_1 \ket{e} \bra{1} + {\textrm{h.c.}} \right) 
\end{eqnarray}
with corresponding coupling structure shown in Fig. \ref{fig:1gate}. Given this choice of laser 
pulses, the dark state $\ket{d} = -\omega_1 \ket{0} + \omega_0 \ket{1}$ decouples from the 
dynamics, which in turn implies that the evolution is reduced to a simple Rabi oscillation between 
the bright state $\ket{b} = \omega_0^{\ast} \ket{0} + \omega_1^{\ast} \ket{1}$ and the excited 
state \cite{fleischhauer96}. The Rabi frequency is $\Omega (t)$. It follows that the qubit subspace 
$M(0)$ evolves into $M(t)$ spanned by $\ket{\psi_k (t)} = e^{-i\int_0^{t}
H^{(1)} (t') dt'} \ket{k} = U(t,0) \ket{k}$, $k=0,1$, which undergoes cyclic evolution if the pulse-pair
satisfies $\int_{0}^{\tau} \Omega (t')dt' = \pi$. The evolution is purely geometric
since $\bra{\psi_k (t)} H^{(1)}(t) \ket{\psi_l (t)} = \bra{k} H^{(1)}(t) \ket{l} = 0$ for $t\in [0,\tau]$.
Under the above conditions, the final time evolution operator $U(\tau,0)$ projected onto the
computational space spanned by $\{ \ket{0}, \ket{1} \}$ defines the holonomic one-qubit gate
\begin{eqnarray}
U^{(1)} (C_{{\bf n}}) & = & {\bf n} \cdot \boldsymbol{\sigma} ,
\end{eqnarray}
where ${\bf n}$ is a unit vector in $\mathbb{R}^3$ and $\boldsymbol{\sigma} = 
(\sigma_x,\sigma_y,\sigma_z)$ are the standard Pauli operators acting on $\ket{0}$ and $\ket{1}$. 
By letting $\omega_0 =\sin (\theta /2) e^{i\phi}$ and $\omega_1 = - \cos (\theta /2)$, 
we find ${\bf n} = (\sin \theta \cos \phi , \sin \theta \sin \phi , \cos \theta )$. 
$U^{(1)} (C_{{\bf n}})$ is a universal one-qubit gate. This can be seen explicitly by noting 
that two pairs of laser pulses corresponding to the unit vectors ${\bf n}$ and ${\bf m}$ 
applied sequentially results in
\begin{eqnarray}
U^{(1)} (C) = U^{(1)} (C_{{\bf m}}) U^{(1)} (C_{{\bf n}}) = 
{\bf n} \cdot {\bf m}  - i \boldsymbol{\sigma} \cdot ({\bf n} \times {\bf m}) .
\end{eqnarray}
This is an SU(2) transformation corresponding to a rotation of the qubit by an angle 
$2\arccos \left( {\bf n} \cdot {\bf m} \right)$ around the normal of the plane spanned 
by ${\bf n}$ and ${\bf m}$. Here, $C_{{\bf n}}$ and $C_{{\bf m}}$ are loops based at 
$M(0)$ and $C = C_{{\bf m}} \circ C_{{\bf n}}$. By suitable choices of ${\bf n}$ and 
${\bf m}$, any desired one-qubit gate can be realized. For instance, the choice ${\bf n} = 
(\cos \phi,\sin \phi,0)$ and ${\bf m} = (\cos \phi',\sin \phi',0)$ results in the phase shift 
gate  $\ket{k} \mapsto e^{2ik \left( \phi' - \phi \right)} \ket{k}$, $k=0,1$, up to an unimportant 
overall phase factor. A Hadamard gate $\ket{k} \mapsto \frac{1}{\sqrt{2}} [(-1)^k \ket{k} + 
\ket{k \oplus 1}]$, $k=0,1$, can be implemented by a single pulse with ${\bf n} = 
\frac{1}{\sqrt{2}}(1,0,1)$.

\subsection{Two-qubit gate}
To complete the universal set, we propose a physical realization of a non-adiabatic holonomic
two-qubit gate in an ion trap setup. Our scheme is a non-adiabatic version of \cite{duan01},
which utilizes the S{\o}rensen-M{\o}lmer setting \cite{sorenson99} to design a holonomic
two-qubit gate. The system consists of an array of trapped ions, each of which exhibiting an
internal three-level structure $0,1$ and $e$. The transitions $0\leftrightarrow e$ and
$1\leftrightarrow e$ for an ion pair in the array is addressed by lasers with detunings
$\pm \nu \pm \delta$ and $\pm \nu \mp \delta$, respectively, where $\nu$ is a phonon
frequency and $\delta$ is an additional detuning. Off-resonant couplings to the
singly excited states $\ket{e0},\ket{0e},\ket{e1}$ and $\ket{1e}$ can be suppressed by
choosing the Rabi frequencies $|\Omega_0 (t)|$ and $|\Omega_1 (t)|$ smaller than $\nu$
\cite{sorenson99}. In this way, the effective two-ion Hamiltonian in the Lamb-Dicke
regime reads
\begin{eqnarray}
H^{(2)} & = & \frac{\eta^2}{\delta}
\left( |\Omega_0 (t)|^2 \sigma_0 (\phi) \otimes \sigma_0 (\phi)  - 
|\Omega_1 (t)|^2 \sigma_1 (-\phi) \otimes \sigma_1 (-\phi) \right) .
\end{eqnarray}
Here, $\eta$ is the Lamb-Dicke parameter ($\eta^2 \ll 1$), $\sigma_0 (\phi) = e^{i\phi/4}
\ket{e} \bra{0} + {\textrm{h.c.}}$ and $\sigma_1 (-\phi) = e^{-i\phi/4}
\ket{e} \bra{1} + {\textrm{h.c.}}$. Note that due to the non-adiabatic nature of our
gate, the ancilla state $a$ of the original adiabatic scheme in \cite{duan01} is no
longer needed. The phase $\phi$ and ratio $|\Omega_0 (t)|^2 / |\Omega_1 (t)|^2 = 
\tan (\theta /2)$ should be kept constant during each pulse pair. By expanding $\sigma_0 
(\phi)$ and $\sigma_1 (-\phi)$, the Hamiltonian $H^{(2)}$ can be decomposed as 
\begin{eqnarray}
H^{(2)} & = & \frac{\eta^2}{\delta} \sqrt{|\Omega_0 (t)|^4 +
|\Omega_1 (t)|^4} \left( H_0 + H_1 \right) .
\end{eqnarray}
The two terms    
\begin{eqnarray}
H_0 & = & 
\sin \frac{\theta}{2} e^{i\phi /2} \ket{ee} \bra{00} -
\cos \frac{\theta}{2} e^{-i\phi /2} \ket{ee} \bra{11} + {\textrm{h.c.}} , 
\nonumber \\ 
H_1 & = & 
\sin \frac{\theta}{2} \ket{e0} \bra{0e} -
\cos \frac{\theta}{2} \ket{e1} \bra{1e} + {\textrm{h.c.}}  
\end{eqnarray}
commute, which implies that 
\begin{eqnarray}
e^{-i\int_0^{\tau} H^{(2)} (t) dt} = e^{-i\pi H_0} e^{-i\pi H_1} 
\label{eq:2qubitU}
\end{eqnarray}
under the $\pi$ pulse criterion $\frac{\eta^2}{\delta} \int_0^{\tau} \sqrt{|\Omega_0 (t)|^4 + 
|\Omega_1 (t)|^4} dt = \pi$. The second factor $e^{-i\pi H_1}$ on the right-hand side of the 
time evolution operator in Eq. (\ref{eq:2qubitU}) acts trivially on the computational subspace 
spanned by $\{ \ket{00}, \ket{01}, \ket{10}, \ket{11} \}$. Thus, $H^{(2)}$ effectively reduces 
to the $\Lambda$-like Hamiltonian $\frac{\eta^2}{\delta} \sqrt{|\Omega_0 (t)|^4 + 
|\Omega_1 (t)|^4} H_0$ from which the holonomic two-qubit gate
\begin{eqnarray}
U^{(2)}(C_{\bf n}) & = & \cos \theta \ket{00} \bra{00} + \sin \theta e^{-i\phi} \ket{00} \bra{11} 
 \nonumber \\ 
 & & + \sin \theta e^{i\phi} \ket{11} \bra{00} - \cos \theta \ket{11} \bra{11} 
 \nonumber \\
 & & + \ket{01} \bra{01} + \ket{10} \bra{10}
\label{eq:2gate}
\end{eqnarray}
follows by analogy of the single-qubit gate above. The path $C_{\bf n}$, being characterized 
by the unit vector ${\bf n} = ( \sin \theta \cos \phi , \sin \theta \sin \phi , \cos \theta )$ in 
$\mathbb{R}^3$, is traversed in the three dimensional subspace spanned by $\{ \ket{00},\ket{11}, 
\ket{ee} \}$ of the internal degrees of freedom of the ions. For instance, a conditional phase 
shift gate $\ket{kl} \mapsto e^{i k l \pi} \ket{kl}$, $k,l=0,1$, can be implemented by chosing 
$\theta = 0$. Due to its entangling nature, $U^{(2)} (C_{\bf n})$ is universal when assisted by 
one-qubit gates \cite{brenner02}.

\subsection{Robustness to decay}
In practical implementations utilizing atomic or ionic systems, $\ket{0}$ and $\ket{1}$ typically 
correspond to stable ground states, while the excited state $\ket{e}$ is unstable. Since the 
excited state is significantly populated in the non-adiabatic scheme, it is important to check 
its robustness to the error caused by the finite life-time of $e$. To test this, we add decay of 
$e$ to the non-adiabatic holonomic gates. We compare the resulting fidelity with that of the 
corresponding adiabatic gate. As a test case, we choose the one-qubit phase shift gate $\ket{k} 
\rightarrow e^{ik\pi/2} \ket{k}$, $k=0,1$, in the non-adiabatic and adiabatic scenarios. This 
gate can be implemented adiabatically utilizing the $\Lambda$-type system, but now by varying 
the two laser couplings independently so as to remain approximately in an instantaneous dark 
state in the limit of large run-time $T$. We assume that the excited state decays to the auxiliary 
ground state level $\ket{g}$ with rate $\gamma$. We model the decay with the Lindblad equation
\begin{eqnarray}
\dot{\varrho}_t = -i[H^{(1)}(t),\varrho_t] + 2L \varrho_t L^{\dagger} - L^{\dagger} L \varrho_t -
\varrho_t L^{\dagger} L ,
\label{eq:lindblad}
\end{eqnarray}
where $\varrho_t$ is the density operator and $L = \sqrt{\gamma} \ket{g} \bra{e}$. Furthermore,
$H^{(1)}(t) = \Omega \left( t \right) \left( \omega_0 \ket{e}\bra{0} + \omega_1 \ket{e}\bra{1} +
{\textrm{h.c.}} \right)$ and $H^{(1)}(t) = \Omega \left( \omega_1 (t/T) \ket{e} \bra{1} + \omega_a (t/T)
\ket{e} \bra{a} + {\textrm{h.c.}} \right)$ is the Hamiltonian in the non-adiabatic and adiabatic settings, respectively.
In the adiabatic case, note that
the $0$-state is decoupled from the excited state and that the $a$ state is another ancillary
ground state level \cite{duan01}. Two hyperbolic secant $\pi$ pulse pairs are chosen to implement
the non-adiabatic phase shift gate. Explicitly, we choose $\Omega (t) \left( \omega_0 ,
\omega_1 \right) = \beta {\textrm{{sech}}} (\beta t) (-1,1)/\sqrt{2}$ and $\Omega (t-\Delta t)
\left( \omega_0' ,\omega_1' \right) = \beta {\textrm{{sech}}} [ \beta (t-\Delta t)]
(-1,e^{-i\pi/4})/\sqrt{2}$, where $\beta$ is the amplitude of the pulses and $\Delta t$ is the
temporal separation of the two pulse pairs. The ideal adiabatic gate is generated in the
$T \rightarrow \infty$ limit by varying the laser couplings $\omega_1 = \sin (\vartheta /2) e^{i\varphi}$
and $\omega_a = - \cos (\vartheta /2)$ along the loop $(\vartheta,\varphi) = (0,0) \rightarrow
(\frac{\pi}{2},0) \rightarrow (\frac{\pi}{2},\pi) \rightarrow (0,\pi) \rightarrow (0,0)$ at
constant speed.

\begin{figure}[htb!]
\centering
\includegraphics[width=0.32\textwidth]{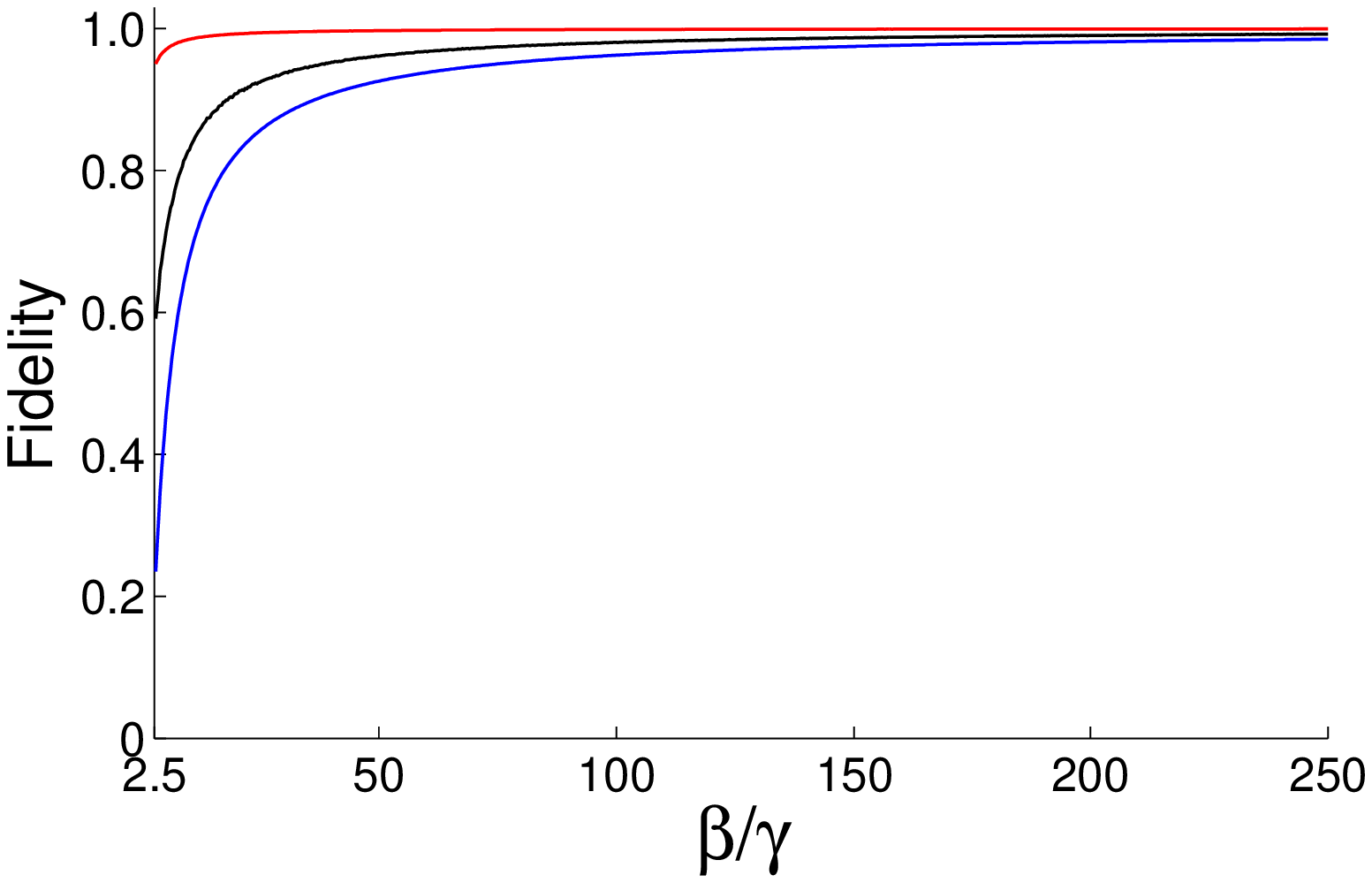}\phantom{.}
\includegraphics[width=0.32\textwidth]{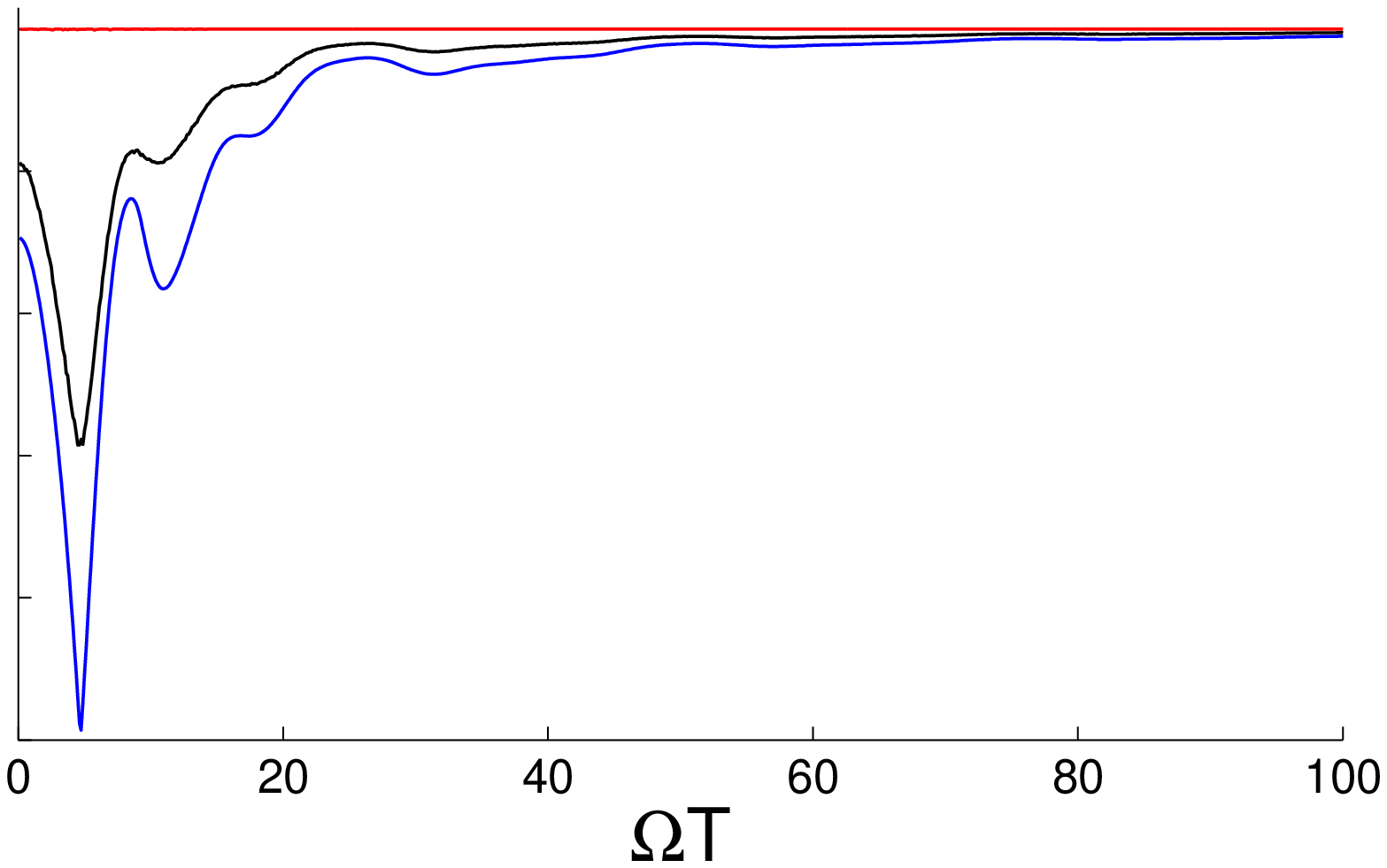}\phantom{.}
\includegraphics[width=0.32\textwidth]{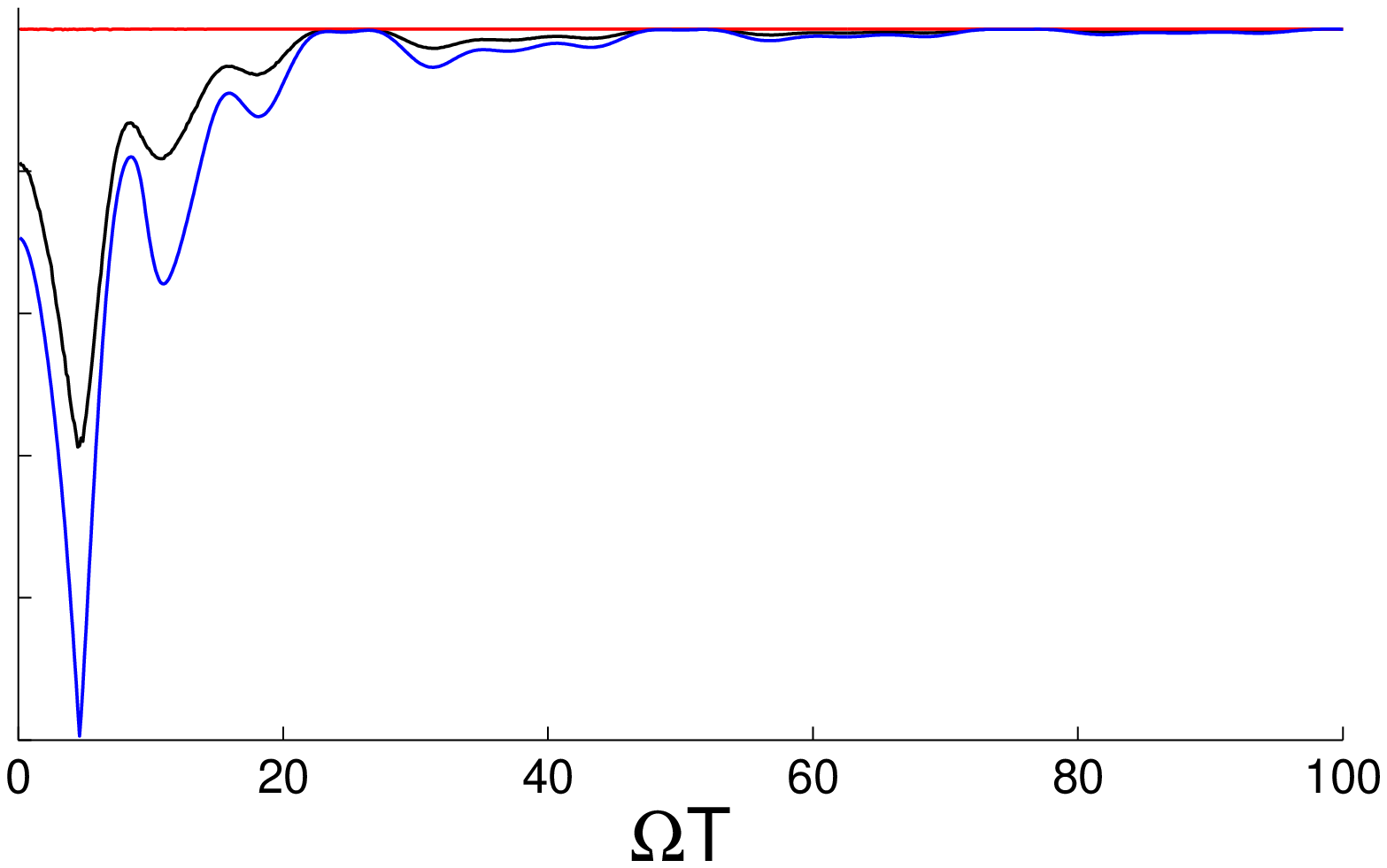}
\caption{Influence of decay with rate $\gamma$ of the excited state $e$ on the non-adiabatic
and adiabatic holonomic phase shift gate $\ket{k} \rightarrow e^{ik\pi /2} \ket{k}$, $k=0,1$.
The effect is quantified in terms of minimum (blue), average (black), and maximum (red) fidelities.
The three panels show from left to right the non-adiabatic gate with decay, the adiabatic gate
with decay, and the adiabatic gate without decay. Choosing hyperbolic secant $\pi$ pulses with
amplitude $\beta$, the non-adiabatic fidelities are shown as functions of the dimensionless
quantity $\beta /\gamma$. We show the adiabatic fidelities as functions of the dimensionless
quantity $\Omega T$, where $\Omega$ is time-independent global strength of laser couplings and
$T$ is the run-time of the gate. We have chosen $\Omega /\gamma = 12.5$ and $\gamma \Delta t = 8$,
where $\Delta t$ is the temporal separation of the two laser pulses in the non-adiabatic setting.
$\Delta t$ is chosen sufficiently large to guarantee negligible pulse overlap for the $\beta /\gamma$
range shown in the left panel.}
\label{fig:graphs}
\end{figure}

In Fig. \ref{fig:graphs}, we show the fidelity $\bra{\xi} U^{\dagger}(C)
\varrho_{{\textrm{out}}} U(C) \ket{\xi}$, computed numerically for $4000$ input states $\ket{\xi}$,
uniformly distributed over the Bloch sphere. Here, $U(C)$ is the non-adiabatic or
adiabatic holonomic gate and $\varrho_{{\textrm{out}}}$ is the output state computed from Eq.
(\ref{eq:lindblad}). The fidelities are shown as functions of the dimensionless quantities
$\beta/\gamma$ and $\Omega T$ in the non-adiabatic and adiabatic cases, respectively. Note that
the pulse duration in the non-adiabatic setting decreases with increasing $\beta /\gamma$ since
the pulse area is set to the fixed value $\pi$. Thus, by increasing $\beta /\gamma$ we effectively
speed up the gate. Furthermore, we have chosen $\Omega /\gamma = 12.5$ and $\gamma \Delta t = 8$,
where the latter choice guarantees that the pulse overlap is negligible for the $\beta /\gamma$
range shown in the figure; a necessary condition to avoid any spurious dynamical contributions
to the gate.

\noindent

The fidelities of the non-adiabatic gate tend monotonically to unity in the large $\beta /\gamma$
limit (left panel). This demonstrates that the non-adiabatic version of the holonomic phase shift
gate can be made robust to decay of the excited state by employing sufficiently short pulses.
A key point with adiabatic holonomic quantum computation is that the population of the decaying
excited state becomes negligible in the adiabatic limit. This behavior is confirmed as the
fidelities of the adiabatic gate in the presence of decay tend to unity in the large $T$ limit
(middle panel). The oscillatory behavior, on the other hand, is due to non-adiabatic effects
originating from the finite run-time of the gate and is thus present also when the decay is set
to zero (right panel). The revivals seen in the fidelities of the adiabatic gate without decay
have been pointed out previously in \cite{florio06,lupo07}.

Other types of errors may affect the gate fidelities. In a separate publication \cite{johansson12}, 
we consider the effect of dephasing (of relevance to superconducting quibts) and different types 
of parameter errors on non-adiabatic and adiabatic holonomic gates. 

\section{Geometric interpretation}
\label{sec:geometry}
To understand the nature of the above holonomic gates, we need to introduce a few 
concepts from differential geometry. A Grassmann manifold $G (N;K)$ is the set of $K$-dimensional 
subspaces of an $N$-dimensional Hilbert space. It is isomorphic to the set of complex $K$-planes 
in $\mathbb{C}^N$. The closed path $C$ of $K$-dimensional subspaces is a loop in $G(N;K)$. The 
set of all bases forms a Stiefel manifold $\mathcal{S} (N;K)$, which is a fiber bundle with $G(N;K)$ 
as base manifold and with the set of $K \times K$ unitary matrices as fibers \cite{bengtsson06}. 
A lift of the loop $C$ in $G (N;K)$ to a loop $\mathcal{C}$ in $\mathcal{S} (N;K)$ corresponds 
to a single-valued choice of gauge. A gauge transformation is a unitary change of bases over 
$C$. The unitary $\boldsymbol{U} (C)$  in Eq. (\ref{eq:general}) is the holonomy matrix associated 
with the loop $C$ in $G(N;K)$.

Now, let us consider the holonomic one- and two-qubit gates described in section 
\ref{sec:implementation}. These gates are associated with loops in $G(3;2)$, where the 
Hilbert spaces relevant for the holonomies are spanned by $\{ \ket{0},\ket{1},\ket{e} \}$ and 
$\{ \ket{00},\ket{11},\ket{ee} \}$ in the one- and two-qubit cases, respectively. We lift the loop 
$C_{\bf n}$ in $G(3;2)$ to a loop $\mathcal{C}_{\bf n}$ in $\mathcal{S}(3;2)$. As noted above, 
each such lift corresponds to a choice of gauge. In the one-qubit case, the loop $C_{\bf n}$ may 
be represented by a set of complex 2-planes spanned by the single-valued vectors
\begin{eqnarray}
\ket{\zeta_1 (t)} & = & U(t,0) \ket{d} = \ket{d} ,
\nonumber \\
\ket{\zeta_2 (t)} & = & e^{i\delta (t)} U(t,0) \ket{b} = 
e^{i\delta (t)} \left[ \cos \delta (t) \ket{b} - i \sin \delta (t) \ket{e} \right]
\label{eq:gaugechoice}
\end{eqnarray}
in the three-dimensional complex vector space $\mathbb{C}^3$. Here, $\delta (t) =
\int_0^t \Omega (t') dt'$ and the global phase factor $e^{i\delta (t)}$ has been
inserted to ensure that $\ket{\zeta_2 (\tau)} = \ket{\zeta_2 (0)}$. The same expressions 
for $\ket{\zeta_1 (t)}$ and $\ket{\zeta_2 (t)}$ apply to the two-qubit case by making the 
replacements $\ket{d} \rightarrow \cos (\theta/2) \ket{00} + \sin (\theta/2) e^{i\phi} 
\ket{11}$, $\ket{b} \rightarrow \sin (\theta/2) e^{-i\phi} \ket{00} - \cos (\theta/2)  
\ket{11}$ and $\ket{e} \rightarrow \ket{ee}$. The loop $C_{\bf n}$ can be visualized 
by noting that $\ket{\zeta_1}$ points in a fixed direction in $\mathbb{C}^3$ around 
which $\ket{\zeta_2}$ rotates. Physically, $\ket{\zeta_1}$ 
represents the dark state and $\ket{\zeta_2}$ describes Rabi oscillations between the bright 
and excited states \cite{fleischhauer96}. The oscillations correspond
to a loop $\mathcal{C}_{\bf n}$ in $\mathcal{S} (3;2)$ represented by the single-valued
gauge choice in Eq. (\ref{eq:gaugechoice}) that projects onto the loop $C_{\bf n}$ of
complex 2-planes in $G(3;2)$. The connection one-form associated with this gauge reads
\begin{eqnarray}
\boldsymbol{\mathcal{A}} = \left( \begin{array}{cc}
 0 & 0 \\
 0 & \Omega (t) dt
\end{array} \right) ,
\end{eqnarray}
which results in the holonomy matrix
\begin{eqnarray}
\boldsymbol{U} (C_{\bf n}) = \boldsymbol{Z} = \left( \begin{array}{cc}
 1 & 0 \\
 0 & -1
\end{array} \right)
\end{eqnarray}
for a single Rabi oscillation. Note that the matrix $\boldsymbol{Z}$ is diagonal in the 
dark-bright basis but in general off-diagonal in the computational basis. An explicit 
calculation confirms that $\sum_{k,l} \boldsymbol{Z}_{kl} \ket{\zeta_k (0)} \bra{\zeta_l (0)} =
 {\bf n} \cdot \boldsymbol{\sigma}$. Similarly, for the composite path $C =
C_{{\bf m}} \circ C_{{\bf n}}$, we obtain $\sum_{k,l} \boldsymbol{U}_{kl} (C)
\ket{\zeta_k (0)} \bra{\zeta_l (0)} = {\bf n} \cdot {\bf m} - i \boldsymbol{\sigma}
\cdot ({\bf n} \times {\bf m})$ from the holonomy matrix
\begin{eqnarray}
\boldsymbol{U} (C) = \boldsymbol{W}^{\dagger} \boldsymbol{Z} \boldsymbol{W} \boldsymbol{Z} .
\label{eq:wz}
\end{eqnarray}
Here, the unitary overlap matrix with components $\boldsymbol{W}_{kl} = \bra{\zeta_k'(0)} \zeta_l(0)
\rangle$ corresponds to an integration of a pure gauge connection one-form $i\boldsymbol{V}^{\dagger}
d \boldsymbol{V}$ along any path $\mathcal{D}$ in $\mathcal{S} (3;2)$ that connects the initial bases
$\{ \ket{\zeta_1 (0)},\ket{\zeta_2 (0)} \}$ and $\{ \ket{\zeta_1' (0)},\ket{\zeta_2' (0)} \}$ of
$C_{\bf n}$ and $C_{\bf m}$, respectively \cite{kult06}. In other words, the loop $C = C_{{\bf m}}
\circ C_{{\bf n}}$ in $G(3;2)$ is lifted to the loop $\mathcal{C} = \mathcal{D}^{-1} \circ
\mathcal{C}_{\bf m} \circ \mathcal{D} \circ \mathcal{C}_{\bf n}$ in $\mathcal{S} (3;2)$, where
the four path segments correspond to the four non-commuting factors on the right-hand side of
Eq. (\ref{eq:wz}).

We end this section by some remarks on the geometrical aspects of the idea put forward by 
Zhu and Wang (ZW) \cite{zhu02} to realize non-commuting quantum gates by implementing 
phase shift gates in different bases. To see how this works, consider the one-qubit phase 
shift gates $\ket{k} \rightarrow e^{i(2k-1) \gamma} \ket{k}$, $k=0,1$, and $\ket{\pm} 
\rightarrow e^{\pm i\gamma'} \ket{\pm}$, $\ket{\pm} = \frac{1}{\sqrt{2}} (\ket{0} \pm \ket{1})$, 
where $\gamma$ and $\gamma'$ are the corresponding cyclic phases. These gates are 
non-commuting and can be used to implement any one-qubit transformation by varying 
$\gamma$ and $\gamma'$. The dynamical phase contribution to $\gamma$ and $\gamma'$ 
can be eliminated either by  employing rotating driving fields with fine-tuned parameters 
\cite{zhu02,zhu03,zhu05} or by driving the qubit along geodesics on the Bloch sphere by 
using composite pulses \cite{solinas03b,tian04,ota09}. These techniques result in non-commuting 
gates solely dependent on the non-adiabatic geometric phases of the cyclic states.  

There are several differences between the geometric phase gates in the ZW setting and 
non-adiabatic HQC proposed in the present work. First, the ZW scheme utilizes holonomies 
generated by loops in $G(N;1)$, which is a fundamentally different space than the relevant 
Grassmannian $G(3;2)$ associated with the holonomies in Eq. (\ref{eq:wz}). Secondly, the 
loops that result in non-commuting gates in the ZW scheme are based at different points 
in $G(N;1)$, while in non-adiabatic HQC all gates are based at a single point in $G(3;2)$, 
namely $M(0)$. Finally, while the dynamical phases vanish for all input states in non-adiabatic 
HQC, all input states except the cyclic ones pick up a non-zero dynamical phase in the geometric 
phase version of the ZW scheme. 

\section{Conclusions}
\label{sec:conclusions}
In conclusion, we have developed a non-adiabatic generalization of holonomic quantum computation
(HQC) with the primary purpose to find ways to construct universal sets of robust high-speed geometric
quantum gates. We have demonstrated an explicit realization of a universal set of holonomic one-
and two-qubit gates in non-adiabatic evolution in three-level $\Lambda$ configurations. The scheme
requires coherent control of fewer levels and behaves simpler under decay of the excited state
compared to the holonomic gates proposed for adiabatic evolution in tripod configurations
\cite{duan01,faoro03,solinas03a}. Our gate opens up for the possibility to realize experimentally 
universal quantum computation on short-lived qubits by purely geometric means.

\subsection{Acknowledgments}
This work was supported by the National Research Foundation and the Ministry of Education
(Singapore). D.M.T. acknowledges support from the National Basic Research Program of China
(Grant No. 2009CB929400).

\end{document}